\documentclass[twocolumn]{aastex631}

\usepackage{amsmath}

\usepackage{savesym}
\savesymbol{tablenum}
\usepackage{siunitx}
\restoresymbol{SIX}{tablenum}

\shorttitle{Thermal Processing of Solids Encountering a Young Jovian Core}
\shortauthors{Barnett and  Ciesla}
\graphicspath{{./}{figures/}}

\begin{document}

\title{Thermal Processing of Solids Encountering a Young Jovian Core}

\correspondingauthor{Megan Barnett}
\email{meganbarnett@uchicago.edu}

\author{Megan N. Barnett}
\affiliation{Department of the Geophysical Sciences, University of Chicago}
\author{Fred J. Ciesla}
\affiliation{Department of the Geophysical Sciences, University of Chicago}
\received{August 13, 2021}
\revised{November 23, 2021}
\accepted{December 15, 2021}



\begin{abstract}

Jupiter's enhancement in nitrogen relative to hydrogen when compared to the Sun has been interpreted as evidence that its early formation occurred beyond the N$_{2}$ snowline ($\sim$ 20-40 AU).  However, the rapid growth necessary to form Jupiter before the dissipation of the solar nebula would lead to the forming planet's core reaching very high temperatures ($>$1000 K), which would lead to it warming its surroundings.  Here, we explore the effects of a luminous planetary core on the solids that it ultimately accretes. We find that a critical transition occurs where very hot (rapidly accreting) cores drive off volatiles prior to accretion, while cool cores (slowly accreting)  are able to inherit volatile rich solids.  Given Jupiter's nitrogen enrichment, if it formed beyond the N$_{2}$ snowline, its core could not have accreted solids at a rate above 10$^{-10}$ M$_{\odot}$ yr$^{-1}$.  Our results suggest that either Jupiter formed in more distal regions of the solar nebula, or nitrogen loss was suppressed, either by its incorporation in more refractory carriers or because it was trapped within ices which devolatilized at higher temperatures.

\end{abstract}



\section{Introduction}
The Galileo and JUNO missions have given important insight into the properties of Jupiter and the conditions under which it formed. Specifically, the Galileo mission provided the first in-situ look into the composition of Jupiter’s atmosphere, finding that volatile species such as Ar, Kr, Xe, C, N, and S were uniformly enriched up to four times relative to solar abundances, with the following JUNO mission suggesting the same is true for O \citep{niemann_galileo_1996, folkner_ammonia_1998, owen_low-temperature_1999, atreya_comparison_1999, atreya_composition_2003, li_water_2020}. As Jupiter’s atmosphere is largely expected to be derived directly from the solar nebula, the gas is expected to reflect the solar abundance of elements or be deficient in particular elements as they freeze out beyond molecular snowlines and are accreted into the core \citep[e.g.][]{oberg_effects_2011}. Thus, observed enrichments must be sourced from solids as frozen-out elements can easily be added independently of hydrogen and helium, which remain predominately as gaseous H$_{2}$ and He under all conditions expected within a protoplanetary disk. However, to be uniformly enriched in all other elements, solids must have formed at much colder temperatures than expected at Jupiter’s current distance from the Sun ($\sim$5.2 AU) to contain all of the volatile species at their observed abundances \citep[e.g.][]{pollack_formation_1996, owen_low-temperature_1999}. 

One possible way to reconcile these observations, suggested by \citet{owen_low-temperature_1999}, would be if Jupiter’s formation was initiated far beyond its current orbital location, where temperatures were low enough to support solids with a nearly solar composition (all elements present at their solar abundance except H and He). The planet would then migrate inwards over time as a result of torques that arise from gravitational interactions with the disk \citep[e.g.][]{nelson_migration_2000, alibert_modeling_2005, paardekooper_giant_2018}. In fact, it is worth noting that the migration of Jupiter from further distances than where it is found now is consistent with the capture of the Jupiter Trojans and preservation of their high inclinations \citep[][]{pirani_consequences_2019}. 

It is this framework that allowed \citet{bosman_jupiter_2019} and \citet{oberg_jupiters_2019} to argue that Jupiter, or at least its core, formed beyond the solar nebula’s N$_{2}$ snow line which is estimated to have been tens of astronomical units from the Sun.  As nitrogen is among the most volatile elements, including even the noble gases \citep{oberg_jupiters_2019}, essentially all other elements would be frozen out beyond the N$_{2}$ snowline, leaving solids in this region with a solar mix of elements.  Accretion of these solids would then result in uniform elemental enrichment in the planet. 

Formation of giant planets at these extreme distances from the Sun is difficult within traditional core accretion models as the 
planetary growth timescale would exceed the typical lifetime of protoplanetary disks \citep[e.g.][]{pollack_formation_1996, hubickyj_accretion_2005}.  More rapid formation, however, is possible in the context of pebble accretion \citep[e.g.][]{lambrechts_rapid_2012, levison_growing_2015} where solid mass is delivered by small solids whose dynamics are controlled largely by their interactions with the gas.  Accretion of these small solids can lead to the rapid production of massive cores in protoplanetary disks, initiating planet formation very early in disk history. In fact, models have shown that a 10 M$_{\earth}$ solid core can grow at 100 AU in less than 1 Myr, and even faster at the shorter distances where the the N$_{2}$ snowline is expected to reside \citep{lambrechts_rapid_2012}. 

As giant planets require rapid growth, it is important to consider the energy balance that occurs during accretion, as planets will get hot and radiate heat to the surrounding environment. Models of Jupiter's formation suggest that the planet reached luminosities exceeding 10$^{-7}$ L$_{\odot}$ throughout its growth and may have reached 10$^{-4}$ L$_{\odot}$ at times \citep{dangelo_growth_2021}.
Radiation escaping from the growing planet may have a significant effect on the local disk environment. For example, \citet{cleeves_indirect_2015} showed that an accreting gas giant could release enough energy to volatilize ices in the area around their orbits, offering a means of detecting these planets in a disk.

The situation considered by \citet{cleeves_indirect_2015} focused on the late-stage growth of a Jupiter-mass planet that had already opened a gap in the surrounding protoplanetary disk.  However, a rapidly growing core early in its evolution may also release enough energy to heat its surface by thousands of Kelvin. In fact, temperatures at the surfaces of pebble-accreting cores can be sufficient to vaporize silicates before they reach the surface \citep[e.g.][]{johansen_pebble_2021}.  In the case of Jupiter forming far from the Sun, if the core was too luminous, nitrogen ice may have devolatilized prior to accretion, preventing the growing core from incorporating this element despite forming beyond the N$_{2}$ snow line. 

As such, in this work we investigate the effect of an accreting giant planet core on the solid material it encounters in its protoplanetary disk and implications for volatile enrichments that could occur during this stage of growth. The next section (Section 2) details the modeling framework used to track the dynamical, thermal, and chemical evolution of pebbles approaching such a core. Section 3 presents the complete histories for particles encountering the core at various stages throughout its growth. We discuss significant trends in our findings in Section 4 and our conclusions are outlined in Section 5 along with discussion for the implication for volatile accretion by growing giant planets.

\section{Methods}
In our model, we simulate a protoplanetary disk with a young planetary core on a circular orbit around a solar-mass star. The core has low enough mass ($\leq$5M$_{\oplus}$) such that it has not opened a gap in the disk and is fully embedded in the gaseous disk.  Solid particles drift inwards from the outer regions of the disk under the influence of gas drag, with some encountering the growing core.  Not all encounters are equal, however, and depend on the details of the core and particle trajectory. To investigate the effect on the nitrogen inventories of accreting solids, we simulate this dynamical evolution, the corresponding thermal evolution, and the resulting chemical evolution of these bodies.

\subsection{Dynamical Evolution}

We simulate the dynamical evolution of solids of various sizes in a protoplanetary disk as they move under the combined gravitational effects of the star and growing planetary core using a method similar to \citet{tanigawa_accretion_2014}. That is, we consider a Cartesian coordinate frame that is centered on and co-rotating with the planet.  The x-axis is defined by the line connecting the core and the star, while the y-axis is oriented in the direction of motion of the planet. The corresponding equations of motion for the particles (focusing on the disk midplane and ignoring vertical motions) are given by:
\begin{multline}
    \ddot{x} = -(\frac{GM_{\odot}(x+a)}{\sqrt{(x+a)^{2}+y^{2}}^{3}} + \frac{GM_{core}x}{\sqrt{x^{2}+y^{2}}^{3}}) + 2\Omega_{0}\dot{y} \\ + \Omega_{0}^{2}(x+a)  -\frac{C_{d}\rho_{g}\pi r_{p}^{2}\Delta v_{p,g}(\vec{v_{px}} - \vec{v}_{gx})}{2m_{p}}
\end{multline}
\begin{multline}
    \ddot{y} = -(\frac{GM_{\odot}}{\sqrt{(x+a)^{2}+y^{2}}^{3}} + \frac{GM_{core}}{\sqrt{x^{2}+y^{2}}^{3}})y - 2\Omega_{0}\dot{x} \\ + \Omega_{0}^{2}y -\frac{C_{d}\rho_{g}\pi r_{p}^{2}\Delta v_{p,g}(\vec{v_{py}} - \vec{v}_{gy})}{2m_{p}}
\end{multline}

where $G$ is the gravitational constant, $M_{\odot}$ is stellar mass, $M_{core}$ is core mass, $m_{p}$ is the mass of the particle, $\rho_{g}$ is the surrounding gas density, $\Omega_{0}$ is the orbital frequency of the core around the central star, $C_{d}$ is the Epstein drag coefficient, $\vec{v_{p}}$ and $\vec{v}_{g}$ are the velocities of the particles and gas respectively, and $\Delta v_{p,g}$ is the magnitude of $\vec{v_{p}} - \vec{v}_{g}$. The last term of each equation represents the acceleration from gas drag \citep{tanigawa_accretion_2014}. Note that while we perform the calculations in the co-rotating reference frame, all figures displayed in this paper are presented in a reference frame centered on the Sun for ease of analysis and interpretation.

The particles in our models are initially defined by their Stokes numbers, St, where:
\begin{equation}
    St = \frac{r_{p}\rho_{p}v_{th,i}}{\rho_{g,i}\Omega_{i}}
\end{equation}
and is used as a proxy for the radius (Stokes numbers change as they migrate into new environments). Particles are assumed to be primarily icy, with a density ($\rho_{p}$) of 1000 kg m$^{-3}$. For this work we consider initial St = 0.01, 0.1, 1.0, or 10.0, but generalize our results to other sizes further below.

Our model focuses on particle movement at the disk midplane and ignores vertical motions. As we ignore the effects of turbulence here, we expect most particles to be located around the disk midplane; the settling time for particles considered here are $\sim$100-10,000 years, which is a short time compared to the lifetime of the disk.  If particles were not fully settled by the time they approach the growing core, it is possible that they could escape an encounter, and avoid accretion.  As our focus is on those particles that are ultimately accreted by the growing core, and accretion requires the particles to be near the plane around the midplane, ignoring  particles at higher altitudes will not change our conclusions.

The physical structure of the protoplanetary disk is taken from \citet{oberg_jupiters_2019}:
\begin{equation}
    \Sigma_{r} = 15,000 \left(\frac{r}{1~\mathrm{au}} \right)^{-3/2} ~\mathrm{kg~m^{-2}}
\end{equation}

\begin{equation}
    T_{b} = 140 \left(\frac{r}{2~\mathrm{AU}} \right)^{-0.65} ~\mathrm{K}
\end{equation}

where $r$ is the distance from the star. Midplane gas densities are then found from:
\begin{equation}
    \rho_{g} = \frac{\Sigma_{r}}{\sqrt{2\pi}H}
\end{equation}
where $H$ is the isothermal scale height:
\begin{equation}
    H = \frac{c_{s}}{\Omega_{K}}
\end{equation}
where $\Omega_{K}$ is the local Keplerian frequency and $c_{s}$ is the sound speed.

The sound speed is calculated using the equation:

\begin{equation}
    c_{s} = \left(kT/\mu m_{H} \right)^{1/2}
\end{equation}

where k is the Boltzmann constant, T is the temperature, m$_{H}$ is the mass of hydrogen, and $\mu$ is the mean molecular weight of the protoplanetary gas which we take to be 2.3.

We consider core masses ranging from 0.5-5 M$_{\oplus}$.  The radius of the solid core is found from the scaling laws for planetary mass-radius developed in \citet{valencia_internal_2006} for Super Earths (1-10 M$_{\oplus}$):
\begin{equation}
    R_{core} = R_{\earth}(M_{core}/M_{\earth})^{0.27}
\end{equation}

In addition to the solid component, a growing core will gravitationally attract surrounding gas in the disk to form a planetary envelope. The radius of this planetary envelope is given by \citet{chambers_steamworlds_2017}:
\begin{equation}
    r_{env} = min[\frac{r_{H}}{4}, r_{B}]
\end{equation}
Here, $r_{H}$ is the core's Hill radius ($r_{H}$ = $a$($\frac{M_{core}}{3M_{\sun}})^{1/3}$), and $r_{B}$ is the core's Bondi radius ($r_B$ = $\frac{GM_{core}}{c_{s}^{2}}$).
For low mass cores, the Bondi radius sets the outer extent of the envelope as it defines the location where gas becomes gravitationally bound to the core.  As the Bondi radius grows at larger masses, however, the differential rotation of the disk becomes important in setting the core's envelope boundary. \citet{lissauer_models_2009} used 3D hydrodynamic simulations to show that gas beyond 25$\%$ of $r_{H}$ will be sheared away from the planet due to this rotation, thus setting this location as the distance beyond which gas no longer remains bound to the core. At the orbital separations considered here, the planetary envelope boundary is defined by the Bondi radius for core masses 0.5-4 M$_{\oplus}$ and is 25$\%$ of the Hill radius for a 5 M$_{\oplus}$ core. Any particles that cross into this envelope-disk boundary are assumed to be accreted, delivering any volatiles they contain to the growing core.

We note that volatile delivery may be affected by planetary envelope recycling flows, as are found in planetary envelope hydrodynamic simulations \citep[e.g.][]{lambrechts_reduced_2017, kurokawa_suppression_2018, popovas_pebble_2018, johansen_pebble_2021}. In these simulations, gas flows within the planetary envelope hinder gaseous volatile delivery to the core as the desorbed volatiles may flow back out to the protoplanetary disk. As such, our results represent an upper limit on volatile delivery to the core.

\subsection{Thermal Evolution}

The temperatures that the particles reach in the disk will be set by the background environment through which they move and radiative heating from the growing core. The contribution from the core is set by the core's luminosity and the distance between the core and the particle. The core surface temperature is defined as:
\begin{equation}
    L_{\text{core}} = 4\pi\sigma R_{\text{core}}^{2}T_{\text{eff}}^{4}
\end{equation}
Ultimately, the luminosity of the growing core is set by the rate of mass accretion which varies over the lifetime of the core. To consider a plausible range of values, we define the luminosity by setting the surface temperature of the solid core to a value between 1000-3000 K; these values correspond to mass accretion rates of $\SI{1.5e-11}{M_{\odot}yr^{-1}}$ to $\SI{1.19e-9}{M_{\odot}yr^{-1}}$ and luminosities of $\SI{7.5e-8}{L_{\odot}}$ to $\SI{6.13e-6}{L_{\odot}}$ for a 1 M$_{\oplus}$ core, all within the range expected from detailed models of Jupiter's growth \citep[e.g.][]{dangelo_growth_2021}.  

As the envelope is assumed to be in hydrostatic balance \citep{rafikov_atmospheres_2006, hori_gas_2011,lambrechts_separating_2014, venturini_critical_2015, chambers_steamworlds_2017}, the rate of energy transfer is constant throughout the envelope, meaning the amount of energy passing into the surrounding nebula at the envelope boundary is equal to $L_{\text{core}}$. The physical structure of the envelope interior is complex, containing both a convective inner layer and radiative outer layer with varying optical depths \citep[][]{hori_gas_2011, venturini_critical_2015, chambers_steamworlds_2017}. However, as we are only concerned with the blackbody temperature experienced by solids at the envelope boundary and we can consider the envelope in steady state, the interior temperature/pressure profiles of the planetary envelope are not necessary for us to calculate here. As such, we follow \citet{rafikov_atmospheres_2006} and \citet{lambrechts_separating_2014} by adopting the optically thin equation to calculate the solid blackbody temperature exterior to the envelope boundary.  

We calculate the blackbody temperature of solids encountering the core using the equation \footnote{We also considered the effect of the optical depth in the disk by having the flux from the core decrease as it passed through the gas, but found this effect to be minor for the cases of interest}:

\begin{equation}
    T_{\text{BB}} = \left(\frac{L_{\text{core}}}{16\pi\sigma r^{2}}+ T_{b}^{4} \right)^{\frac{1}{4}}
\end{equation}

This treatment assumes that the solid instantaneously equilibrates with the radiation field of the core and ignores the diffusion of heat into the interior of the solid. However, given that we expect that the resulting desorption of molecules will occur from the surface of the solids, such an assumption is justified.

\begin{figure}[ht]
    \centering
    \includegraphics[width=8.75cm]{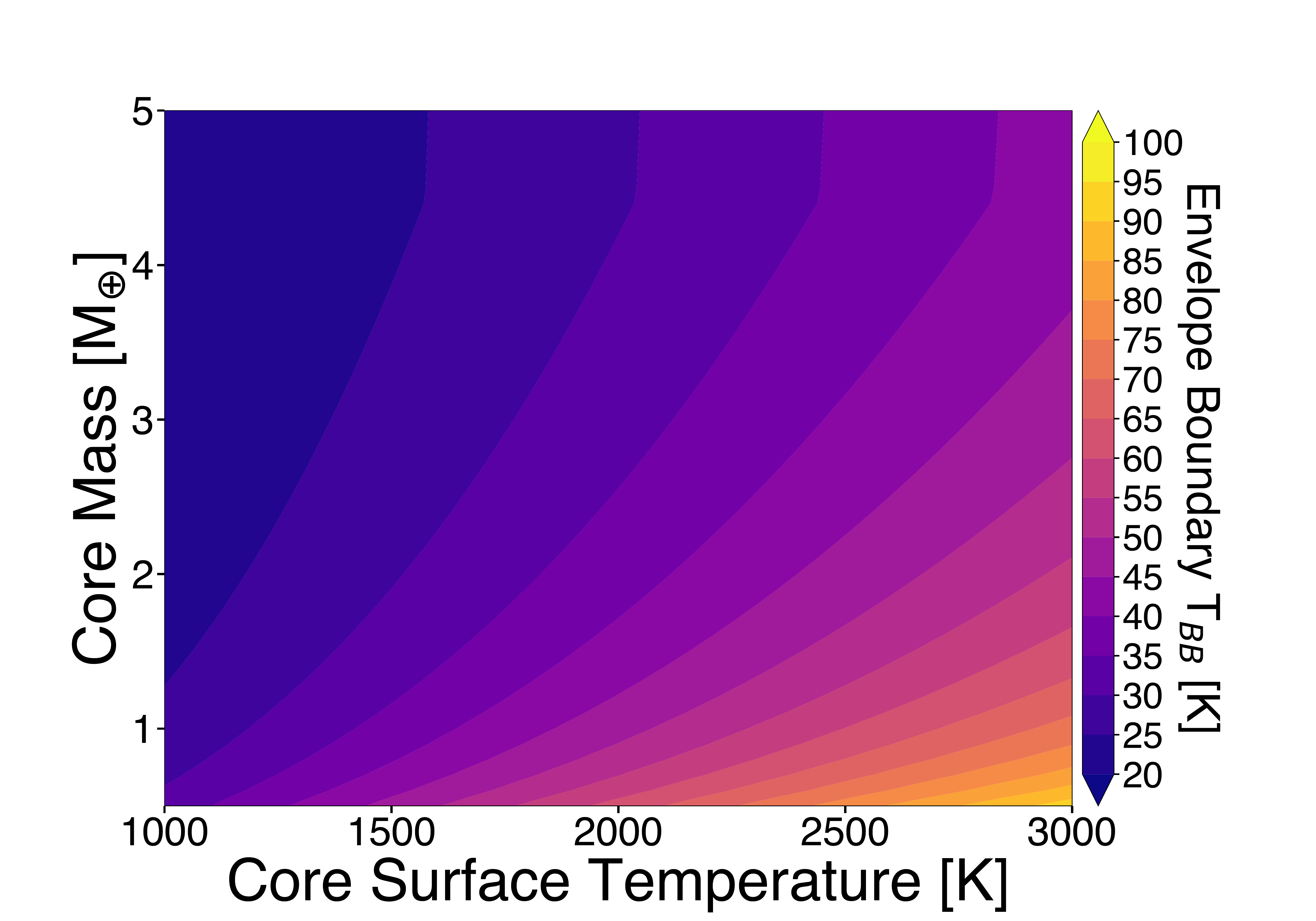}
    \caption{Blackbody temperature experienced by solids at the envelope boundary for the entire range of simulated core masses and core surface temperatures. Lower mass cores have smaller planetary envelope radii, resulting in solids coming in closer proximity to the core before accretion. Due to this closer proximity, solids experience higher blackbody temperatures before accretion. Core masses of 0.5 M$_{\oplus}$ lead to the highest solid blackbody temperatures at the envelope boundary for each simulated core surface temperature, while 5 M$_{\oplus}$ core masses result in the lowest solid blackbody temperatures.}
    \label{fig:fig3.5}
\end{figure}

Blackbody temperatures for solids at each core mass envelope boundary for the range of core surface temperatures are displayed in Figure \ref{fig:fig3.5}. For the range of core masses and core surface temperatures we consider, the resulting temperature enhancements felt by solids at the various cores' envelope boundaries range from $\sim$2K - 100K. The radius of the planetary core's envelope scales with core mass, which results in radiation emitted from the surface of larger cores having to travel farther to reach the respective envelope boundaries. This longer distance traveled results in lower blackbody temperatures experienced by solids at the core boundary of larger mass planets.

\subsection{Chemical Evolution}

As particles are warmed in the disk, species in their ice mantles may begin to volatilize.  This occurs via thermal desorption at a rate described by the Polanyi-Wagner relation \citep[e.g.][]{hollenbach_water_2008, piso_co_2015}:

\begin{equation}
    k_{desorption,i} = \nu_{i} e^{\frac{E_i}{T_{BB}}} [\mathrm{s}^{-1}]
    \label{eq:pw}
\end{equation}

Where $\nu_{i}$ = $\SI{1.6e11}{}\sqrt{(E_{i}/k)(m_{H}/m_{i})}$ s$^{-1}$ is the vibrational frequency of a given species and $E_{i}$ is its corresponding binding energy. As our focus here is on the nitrogen content of the solids, we consider the two most abundant carriers, N$_{2}$ and NH$_{3}$  \citep{oberg_spitzer_2011, pontoppidan_nitrogen_2019}. N$_{2}$ is  more volatile with a binding energy of 1050K  \citep[$\sim$15-25K freeze-out temperature,][]{bisschop_desorption_2006, fayolle_n2and_2016}, while NH$_{3}$ desorbs at higher temperatures with a binding energy of 3800K \citep{oberg_jupiters_2019}. We set the initial abundances relative to hydrogen of N$_{2}$ and NH$_{3}$ as $\SI{3.e-5}{}$ and $\SI{7.e-6}{}$, respectively \citep{oberg_jupiters_2019}.

In our model, we only consider loss of molecules from the particles' ice mantles; we ignore freeze-out as the particle sizes we consider (Stokes numbers \textgreater 0.01) have sufficient drift velocity such that desorbed molecules are unlikely to freeze back out onto the solid before it has drifted away. That is, freeze-out timescales around the disk midplane are on the order of $\sim$1 year, while solids and gas have relative velocities of $\sim$ 1 m s$^{-1}$ or larger \citep[e.g.][]{weidenschilling_aerodynamics_1977}, allowing the solids to drift away from any desorbed molecules before they freeze-out again. Further, desorbed molecules would likely freeze-out on the smallest solids present (micron-sized fine dust with St $\sim$ 10$^{-4}$) as these particles provide the greatest total surface area.  The desorption of molecules from the ice mantle is calculated using a first order Euler method, using the rate given in Equation (\ref{eq:pw}). We ignore other forms of molecular desorption or destruction (via UV, X-rays, or cosmic rays) as we are focused on regions around the disk midplane where fluxes of energetic particles and photons are expected to be low. 

\section{Results}

In order to sample the range of dynamical encounters between particles and the growing core, we simulated the evolution of 360 different particles for each set of conditions (particle Stokes number, core semi-major axis, core mass, and core surface temperature). These particles began on orbits that were 1-2 AU greater than the semi-major axis of the planet, distributed as a ring around the star and separated by 1 degree from one another.  The particles were then allowed to drift inwards over time due to the effects of gas drag, accounting for the gravitational effects of the growing core and central star. Simulations were run for $\sim$10$^{4}$ years of model time, sufficient enough for particles to either be accreted by the core or drift inside of the core's orbit such that continued encounters would not occur.

\begin{figure}[htb]
    \centering
    \includegraphics[width=8.5cm]{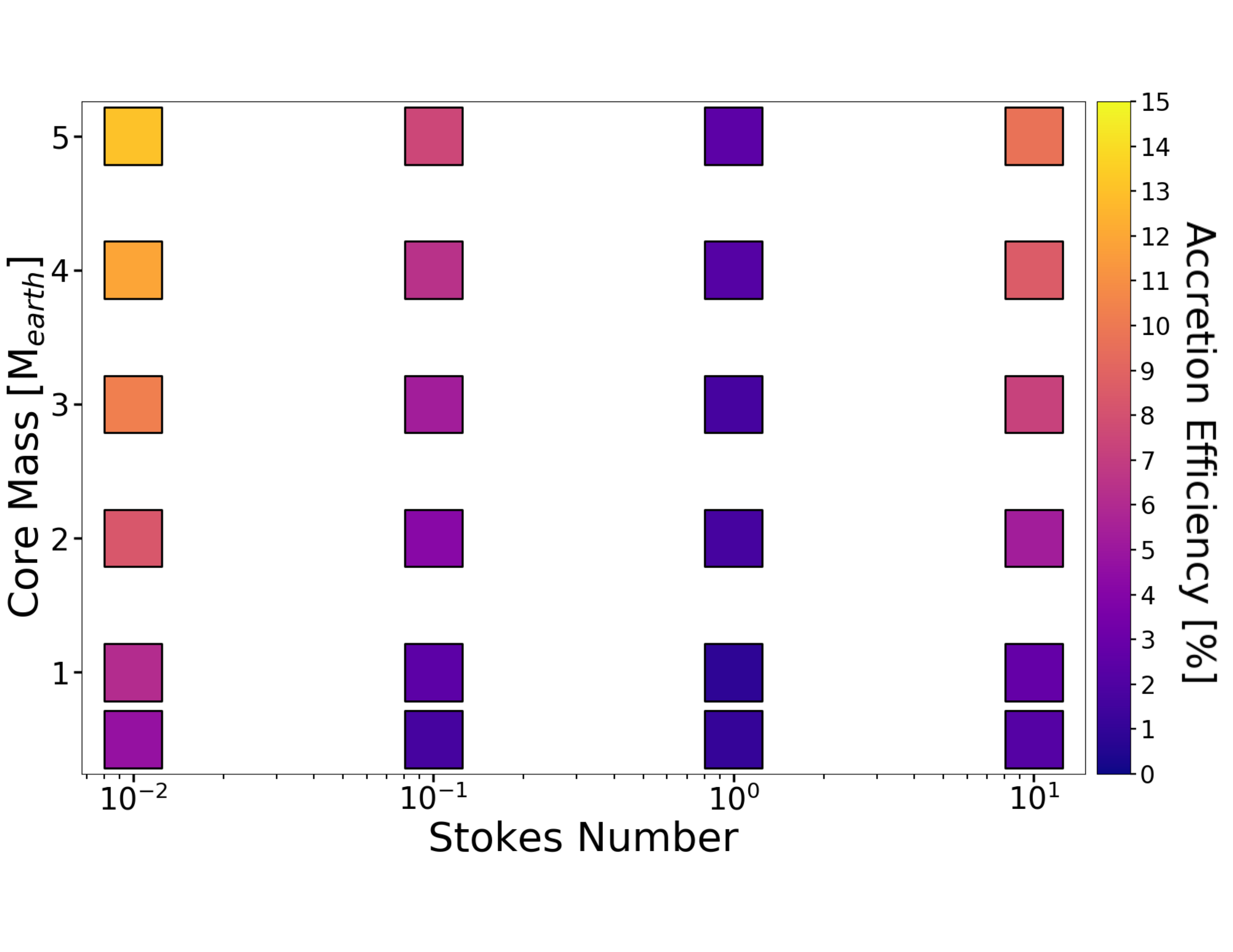}
    \caption{Accretion efficiency for particles initialized with all combinations of model core mass and Stokes number. Core/star orbital separation is fixed at 40AU. Accretion efficiency is highest for smallest Stokes number particles (St = 0.01), followed by the highest Stokes number particles (St = 10). Accretion efficiency is lowest for St = 1 particles.}
    \label{fig:fig3.3}
\end{figure}

In each simulation, only a small fraction ($<$15$\%$) of the drifting particles were accreted by the core.  The fraction of particles accreted, or ``accretion efficiency'', is shown in Figure \ref{fig:fig3.3} as a function of core mass and Stokes number. The lowest accretion efficiency for all cases is seen in St=1 particles where, depending on core mass, 0.8-2.5$\%$ of particles are accreted by the core.  This occurs because these are the most rapidly drifting particles in the disk, migrating through the range of radial distances where they may have a close (accretionary) encounter with the growing core in the least amount of time. Accretion efficiencies increase as one moves away from St=1 particles as smaller solids remain in the vicinity of the core's orbit for longer periods of time, increasing the likelihood of a close encounter. 

The accretion efficiencies found in our model are consistent with trends and averages found in pebble accretion studies from \citet{lambrechts_forming_2014} and 2D pebble accretion models from \citet{ormel_emerging_2017}, but the details of accretion efficiency will vary with initialized particle population and model set-up. Our model simulates the extended dynamical paths of particles which are initialized outside the core's orbit and encounter the core as they drift inwards. This provides a more realistic picture of the geometries of pebble trajectories for pebbles that encounter the core. Other models evaluating accretion efficiency take a different approach, and instead inject pebbles close to the planetary core \citep[e.g.][]{popovas_pebble_2018}. This can lead to different trends in accretion efficiency with particle size, as disparate pebble population locations lead to different encounter geometries.

\begin{figure}[htb]
    \centering
    \includegraphics[width=8.5cm]{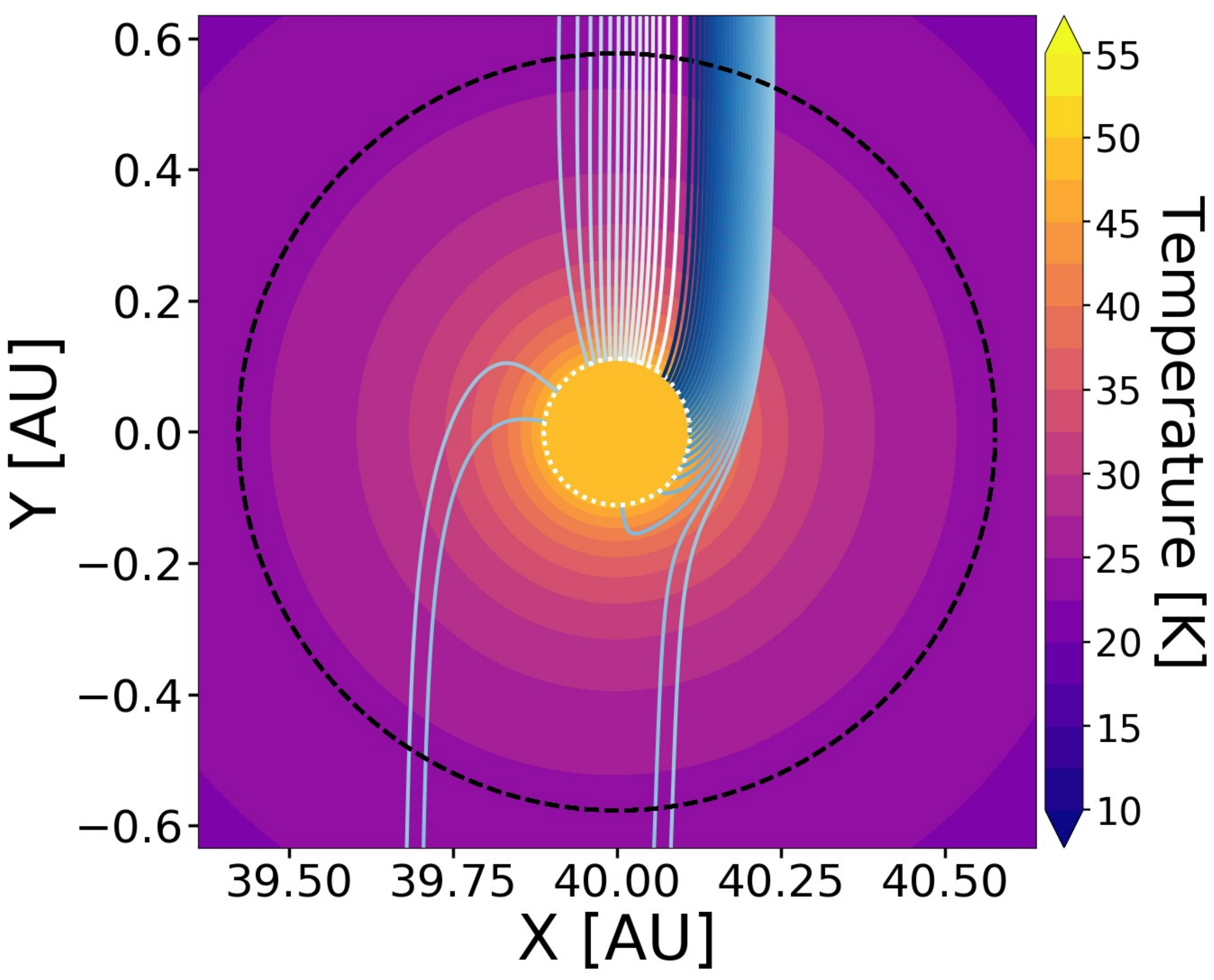}
    \caption{Particle trajectory lines (solid white to blue lines) from particles started at different azimuthal locations over-plotted on the temperature contour map (colorbar) describing the temperature environment of the protoplanetary disk near the planetary core. This model features a 3 M$_{\Earth}$ planetary core located 40AU from the central star with a surface temperature of 3000K. Particle trajectories are for all accreted particles with St = 0.01. Black dashed and white dotted lines mark the Hill radius and envelope boundary, respectively.}
    \label{fig:fig3.1}
\end{figure}

Figure \ref{fig:fig3.1} shows the paths followed by all accreted particles in the St=0.01, $M_{core}$=3M$_{\oplus}$ case for a core orbiting at 40 AU with a surface temperature of 3000K. We display particle trajectories in the area directly adjacent to the core as this is where the particles approach close enough to experience heating, although, again,  we simulate particle dynamics through the entire protoplanetary disk. While all shown particles are eventually accreted, we note that their accretion trajectories fall into two categories: \emph{directly} accreted particles which impact the core soon after crossing its Hill radius, and \emph{indirectly} accreted particles that enter and leave the Hill sphere only to return again before accretion. The enhanced accretional cross section of the core compared to its physical cross-section is due to the gas drag-regulated velocities of the drifting solids that allows pebble accretion to be so efficient as a means of growth \citep[][]{lambrechts_rapid_2012, lambrechts_forming_2014, kretke_challenges_2014, chambers_giant_2014, levison_growing_2015, johansen_forming_2017}.

\begin{figure*}[htb]
    \centering
    \includegraphics[width=\textwidth]{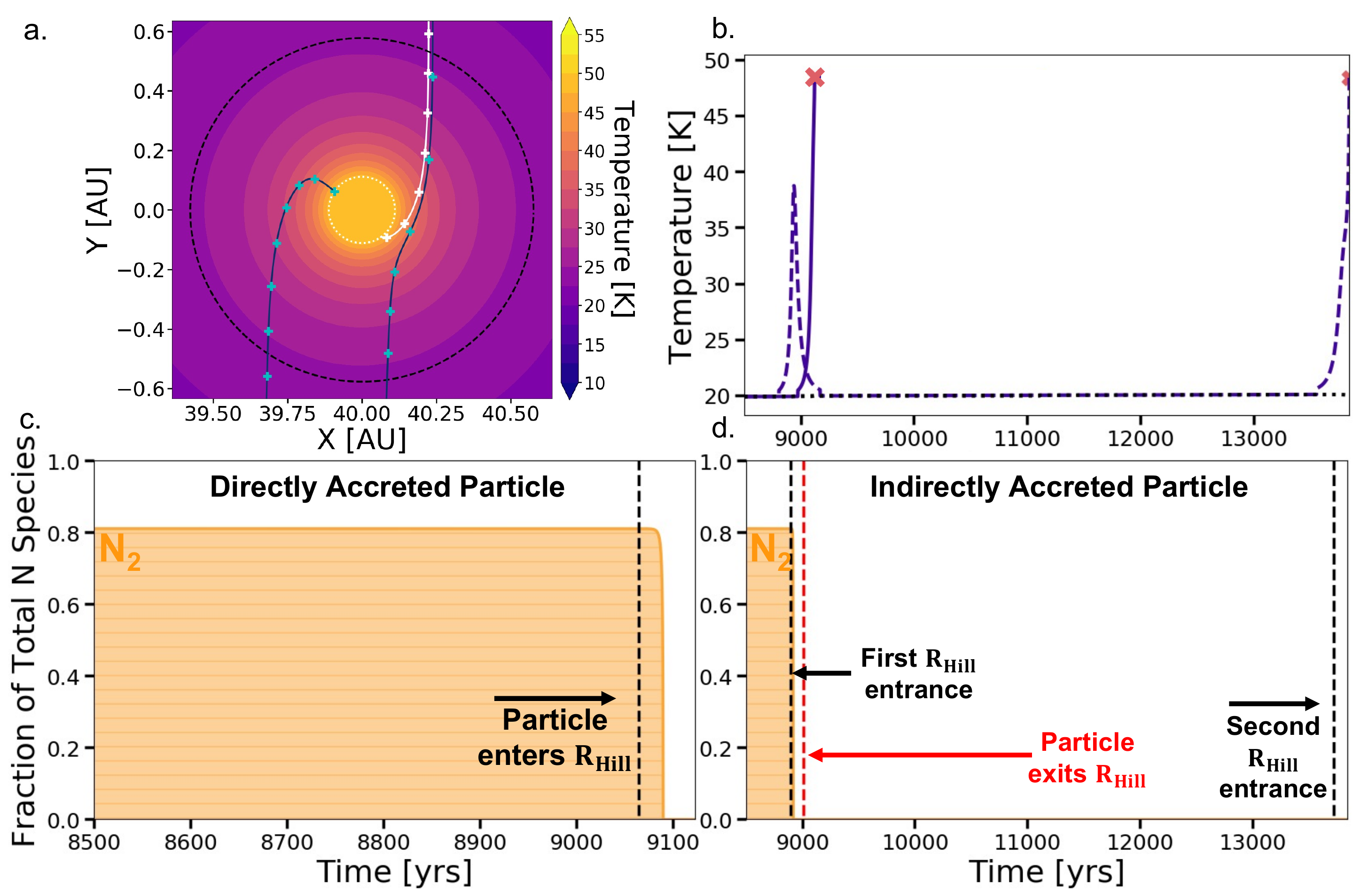}
    \caption{Particle trajectories (a), thermal (b), and chemical evolution (c and d) comparison for particles with St = 0.01 around a 3 M$_{\Earth}$ core at a = 40AU and core surface temperature of 3000K. The indirectly accreted particle (solid blue line in panel a, and dashed line in panel b) and directly accreted particle (solid white line in panel a, solid line in panel b) are identical except for different azimuthal starting locations. We show the thermal and chemical evolution starting from 8500 years into the particles' evolution, as this is when the particles first encounter the core. Crosses (panel a) denote 10 year time-points for the directly accreted particle track (white, 9060-9120 years) or 20 year time-points for the indirectly accreted particle tracks (cyan, 8900-9000 years on right-hand side and 13710-13850 years on left-hand side).  Pink x’s (panel b)  correspond to time of accretion for each particle and represent the endpoints of panel c and d plots.}
    \label{fig:fig3.2}
\end{figure*}

Figure \ref{fig:fig3.2} shows the variation in thermal and chemical evolution that different particles may experience as a result of their close encounter with the core.  Here we show two different particle trajectories (Figure \ref{fig:fig3.2}), one directly accreted particle from Figure \ref{fig:fig3.1} (white line in Figure \ref{fig:fig3.2}a) and one indirectly accreted particle (dark blue line in Figure \ref{fig:fig3.2}a). The directly accreted particle's temperature monotonically increases on its way to being accreted, reaching values near 50K before entering the core's envelope (Figure \ref{fig:fig3.2}b). In this case, the particle's N$_{2}$ ice completely desorbs from the ice mantle outside of the core's envelope but leaves NH$_{3}$ ice, allowing just 20$\%$ of the original nitrogen inventory to be accreted by the core (Figure \ref{fig:fig3.2}c).  

The indirectly accreted particle is similarly warmed during its initial close passage to the core, reaching temperatures of $\sim$40K before moving back outside the core's Hill radius, cooling as it migrates away.  While this temperature is cooler than that reached by the directly accreted particle, it is sufficient to drive off the N$_{2}$ ice in the particle's mantle (Figure \ref{fig:fig3.2}d).  Interestingly, upon exiting the Hill sphere of the embryo, this particle would be available for incorporation into other planetesimals or cores that may be nearby, delivering nitrogen-depleted solids to these growing bodies.  In the absence of such an event, the particle returns to the core and is accreted, also delivering a sub-solar amount of nitrogen to the growing planet. 

\begin{figure*}[ht]
    \centering
    \includegraphics[width=\textwidth]{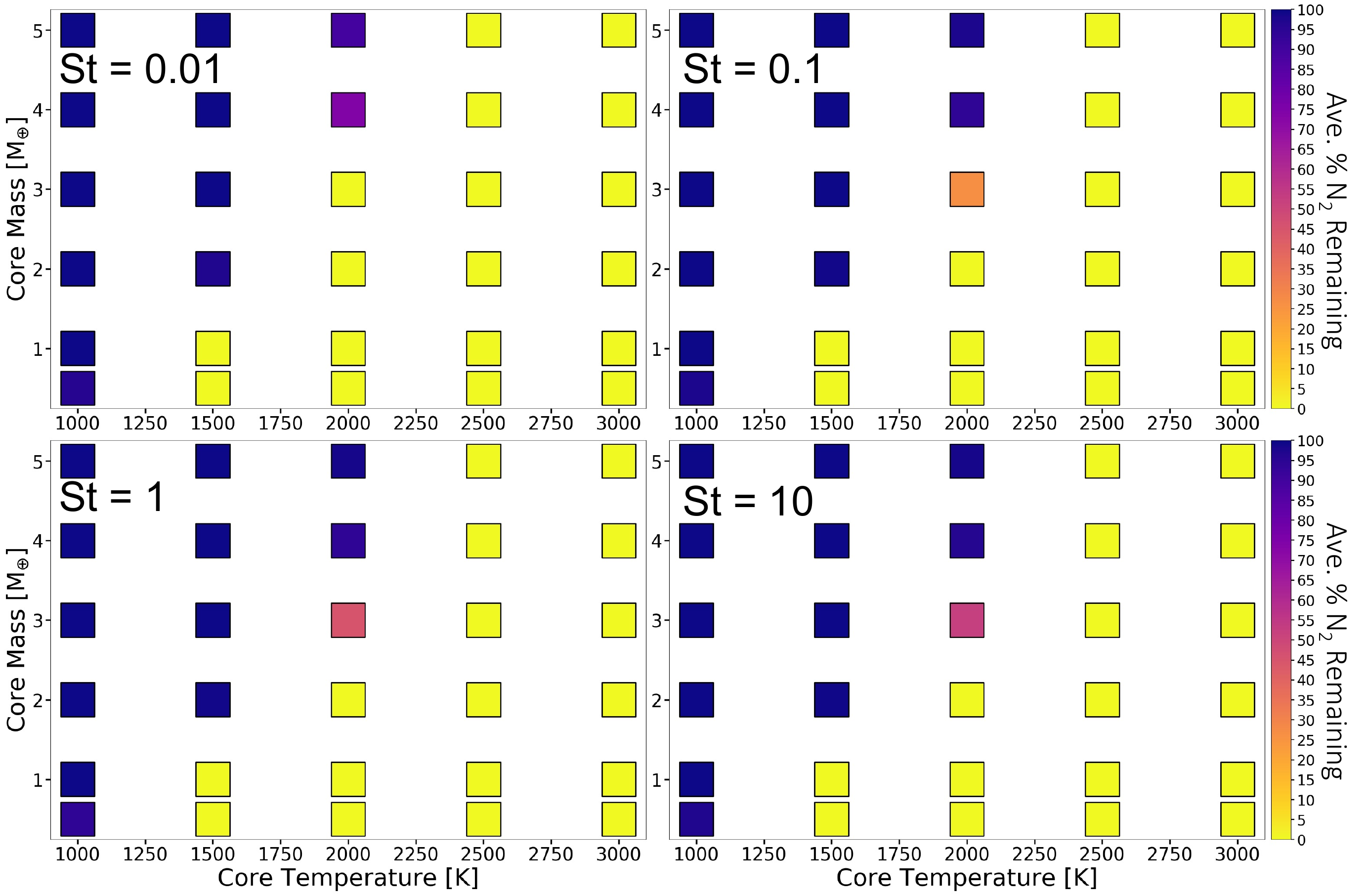}
    \caption{Percent particles with N$_{2}$ as the dominant nitrogen species (colorbar) for models initialized with all combinations of core mass and core temperature. Subplots are results for particles with Stokes numbers 0.01 (upper left), 0.1 (upper right), 1 (lower left), 10 (lower right). The planet was assumed to orbit at 40 AU in these simulations.}
    \label{fig:fig3.4}
\end{figure*}

With our simulated collection of trajectories, thermal histories, and subsequent nitrogen inventory evolution, we can then determine how much of this element is delivered to a growing core for the conditions considered. These results are displayed in Figure \ref{fig:fig3.4}, which shows the average percentage of remaining N$_{2}$ ice on accreted particles for all core mass-temperature and particle Stokes number combinations considered. In all cases, we see that cores with temperatures of 2500K or higher are too luminous for N$_{2}$ ice to be retained; particles experience significant heating before accretion and thus would be unable to enrich a growing core in this element. At temperatures of $\sim$1000K or lower, the low amount of energy radiated from the core does not significantly warm incoming pebbles before accretion, allowing them to retain their full nitrogen inventories.

Intermediate to these temperature regimes, we see a transition occur in the 1500K-2000K runs where the retention of N$_{2}$ ice begins to differ based on the sizes of the particles and the core mass. More specifically, as the core mass decreases, less N$_{2}$ ice is  retained by the accreting particles. This effect arises because both the Bondi radius and Hill radius of the core, and thus the size of its envelope, depend on the core's mass. Particles are considered accreted when they pass the envelope-disk boundary, thus larger envelope radii translate to lower solid blackbody temperatures at the boundary as particles are relatively farther from the core when they are accreted. 

Though this temperature difference can be seemingly minor between core masses (on the order of $\sim$2-10K difference between core masses 2 M$_{\Earth}$ to 3 M$_{\Earth}$), this can lead to significant differences in volatile loss given the exponential dependence of desorption rates on temperature. Additionally, as particles decrease in size (Stokes number), a smaller fraction retains their N$_{2}$ when compared to larger pebbles (see 3 M$_{\Earth}$ core model for all Stokes numbers in Figure \ref{fig:fig3.4}). This is due to two effects: (1) larger particles are less coupled to the gas and thus fall through it at faster rates than the small particles where drag slows down their accretion, and (2) larger particles have larger inventories of ice and thus take longer to lose their nitrogen than the small particles. Both of these effects result in greater volatile loss from smaller particles, a trend that would continue to smaller particles than those considered here.

The dependence of volatile loss on core mass and particle size is best captured in the T$_{core}$ = 2000K runs in Figure \ref{fig:fig3.4}. We see that solids approaching very low mass cores (0.5-2 M$_{\Earth}$) experience full stripping of their frozen N$_{2}$, while higher mass cores (4 and 5 M$_{\Earth}$) fully maintain all N$_{2}$ for all particle Stokes numbers. However, solids encountering the 3 M$_{\Earth}$ core experience progressively more N$_{2}$ stripping as the pebble Stokes number decreases. This transition occurs because the temperature experienced by solids at the 3 M$_{\Earth}$ envelope boundary is fairly moderate at $\sim$ 33K for a core surface temperature of 2000K, thus the amount of time each solid spends in the vicinity of the core before accretion has a significant effect on its N$_{2}$ inventory.

\section{Discussion}

Our results indicate that the surface temperature (luminosity) of a growing core will be the primary factor in determining whether N$_{2}$ ice is retained and delivered during accretion. Additionally, we find that the presence of a planetary envelope is required for volatile delivery as the envelope increases the effective accretionary boundary radius by a factor of $\sim$500-2500. Solids approaching the core are then accreted at much larger radii from the core and as such reach much lower temperatures before accretion than they would otherwise.  In the absence of an extended envelope, significant warming would occur prior to accretion, leading to volatile-poor planets.

Cores with temperatures of $\lesssim$1500K will readily accrete all the nitrogen that its feed stock is able to carry, while cores with temperatures of $\gtrsim$2500K would be depleted in nitrogen. At temperatures between 1500-2000K, the nitrogen story is more nuanced, depending on the size of the core and the particles being accreted. As these second order effects become important at core temperatures of 1500-2000K, we define this as the transition point between when a solar inventory of nitrogen would be delivered to the core versus N-depleted solids, allowing us to constrain the conditions under which a growing giant planet core would inherit a solar mix of this element. Given that the temperature of the core's surface is set by the accretion rate of solids, this transition point sets an upper limit on the accretion rate that would allow the chemical constraints inferred for Jupiter's formation to be met. To maintain a core temperature below this critical value, the mass accretion rate must satisfy:
\begin{equation}
    \dot{M} \lesssim \frac{4 \pi R_{core}^{3} \sigma T_{core}^{4}}{G M_{core}}
\end{equation}
For a 5 M$_{\oplus}$ core, this implies that $\dot{M} \lesssim$ 10$^{-10}$ M$_{\odot}$ yr$^{-1}$ to stay below $T_{core}<$2000 K. This value is in the range estimated for Jupiter's formation ($>$10$^{-11}$ M$_{\odot}$ yr$^{-1}$), but suggests that some of the higher values ($\sim$10$^{-9}$ M$_{\odot}$ yr$^{-1}$) \citep{lambrechts_rapid_2012,lambrechts_forming_2014,dangelo_growth_2021} must be ruled out. Lower mass cores thus have even lower limits on the the critical mass accretion rate needed to maintain frozen N$_{2}$ (Figure \ref{fig:fig3.5}). 

The core surface temperature where cores transition from being able to accrete volatiles at a given temperature versus being unable to do so will depend on the location in the disk where the core forms as the semi-major axis (a) also impacts the extent of the envelope ($r_{H} \propto a$ and $r_{B} \propto a^{0.65}$). Smaller semi-major axes would push the transition point to lower core temperatures as particles approach closer to the core before accretion and would therefore be exposed to higher radiation fluxes and reach higher temperatures. Thus, if Jupiter were to form closer to the Sun, the upper limit on the accretion rate would be even lower. For simulations at 30 AU, similar to those presented here, we found a transition temperature of $\sim$1500K for a 5 M$_{\oplus}$ core, which would limit accretion rates to below $\SI{7.5e-11}{M_{\odot} yr^{-1}}$. 

This finding is important in considering the possibility that Jupiter formed closer to its current orbital separation. In discussing the Galileo probe data, \citet{owen_low-temperature_1999} offered another possible mechanism for Jupiter's atmospheric volatile enhancement, suggesting that temperatures at Jupiter's current orbit were much colder than previously believed. This idea was recently explored by \citet{ohno_jupiters_2021}, who suggested that a dust enhancement just inside of the water snow line could have led to a shadowing of Jupiter's orbit, vastly reducing temperatures there. However, though decreasing the background temperature may allow N$_{2}$ ice to form in this location, forming Jupiter closer to its current orbital separation causes a dramatic increase in solid blackbody temperatures during accretion onto larger ($\geq5M_{\oplus}$) core masses. 

While for a 5M$_{\oplus}$ core with T$_{core}$ = 2000K at 40AU (the transition point for a 5M$_{\oplus}$ core) the solid blackbody temperature is $\sim$29K at the planetary envelope boundary, moving that same core to 5AU increases the solid blackbody temperature to $\sim$78K, which would lead to incredibly rapid N$_{2}$ loss. We find that the transition point for such a core at 5AU now occurs around a temperature of 700K, corresponding to a mass accretion rate of $\sim\SI{2.5e-12}{M_{\odot} yr^{-1}}$ and a core doubling time of $\sim\SI{5e6}{yrs}$. This formation time is likely too slow to form a sufficiently massive core before gas dissipation in the protoplanetary disk occurs, implying that forming Jupiter at 5AU with its atmospheric volatile enhancements would prove challenging if nitrogen is expected to be accreted by higher mass cores. However, as lower mass cores have envelope boundaries set by the Bondi radius, which in turn depends on the disk background temperature, shadowing would allow these lower mass cores to accrete nitrogen ice even at 5AU. While shadowing does potentially provide a mechanism for Jupiter to accrete nitrogen ice at 5AU during the early stages of core accretion (M$_{core}$ \textless 5 M$\oplus$), the specifics of this scenario should be investigated in future work.

\section{Conclusions and Summary}
In this work, we have found that the luminosity of growing cores may be sufficient to drive off volatiles from solids before they are accreted by a forming planet. Thus, even if a given planet formed beyond a snow line in the disk, it may still fail to accrete the particular species that freezes out at that location. In other words, the chemical composition of a planet will not simply reflect the local environment where it formed but instead is set by a complex interplay of its formation location and its accretion history. Planets that form rapidly or experience rapid accretion rates will devolatilize solids to some degree before they are accreted. In the case of Jupiter, the fact that its nitrogen abundance is uniformly enhanced with other volatile elements suggests it formed in a very cold environment and slowly enough to prevent loss of volatiles during accretion.  

We note that elemental enhancements may be possible at higher mass accretion rates under certain circumstances. Here we considered accretion at 40 AU, but accretion further out in the disk, sufficiently beyond the snow line, may be conducive to retention of all volatiles. As discussed above, the extent of the core's envelope is proportional to either $r_{H}$ or $r_{B}$, both of which increase with semi-major axis. This would lead to lower radiative fluxes emerging from the envelope for the same core temperatures considered here. This, combined with the cooler background temperatures at these locations, may allow volatiles to be retained more readily. Additionally, while our model assumes direct desorption of all molecules off the grain as temperatures rise, this may not be the case for nitrogen molecules in the solar nebula. \citet{owen_low-temperature_1999} posited that one mechanism of nitrogen delivery to a forming Jupiter could be through trapping of N$_{2}$ molecules in amorphous water ice. In this case, N$_{2}$ would only be lost when the surrounding water ice desorbs from the grain, but this occurs at much higher temperatures ($\sim$150-180K) than N$_{2}$ desorption. This mechanism may allow delivery of N$_{2}$ at higher core surface temperatures than the upper boundary characterized in our results. Further, the model considered here assumed steady-state (constant) accretion of solids by the core; if accretion is instead episodic, it is possible large amounts of mass could be delivered while the core remains relatively cool. These conditions should be considered in future studies.

Additionally, while this work focuses on evaluating the conditions under which nitrogen can be accreted by Jupiter in the solid phase, another potential mechanism for explaining volatile enhancement in Jupiter’s atmosphere has recently been suggested by \citet{schneider_how_2021-1}.  They propose that volatile enhanced gas generated as nitrogen ice-rich pebbles passed interior to the N$_{2}$ snowline could provide Jupiter’s observed nitrogen enhancement. As the N$_{2}$ would be accreted in the gas phase, there would no longer be a constraint placed on the growth rate of Jupiter's core.

The outcomes described here develop due to the paths that accreting solids take through the radiation field of a growing planet. In our work, we have not accounted for detailed changes in gas flow due to the gravitational influence of the core in its immediate vicinity as has been done in other studies \citep[e.g.][]{okamura_growth_2021}. This could alter the trajectories of the grains and their exposure time to high radiation flux. However, given the exponential dependence of desorption rate on temperatures, this effect is likely minor in determining particle volatile loss, possibly slightly shifting the temperatures where secondary effects become important but not likely to change the general conclusions detailed here.

Giant planet accretional history will also be important when interpreting the elemental ratios observed in exoplanet atmospheres. One of the primary observations to be carried out by the James Webb Space Telescope  will be the determination of C/O ratios in the atmospheres of giant exoplanets. Given that carbon and oxygen are expected to be present across a number of molecular carriers with a wide range of volatilities \citep[e.g.][]{li_earths_2021}, these elements may be driven off to varying degrees depending on their dominant molecular carrier in the protoplanetary disk and accretionary history. This would allow for the planet to inherit elemental ratios that differ significantly from its host star. Such possibilities were motivated by early analyses of WASP 12b observations \citep[e.g.][]{ali-dib_carbon-rich_2014, madhusudhan_carbon-rich_2011, madhusudhan_toward_2014, oberg_effects_2011} which investigated how the formation location or migration of the planet through its protoplanetary disk was related to the planet's atmospheric composition. Based on the results presented here, mass accretion rate will also play a role in setting the composition of a planet and the apparent evolution of a planet relative to its host star, and must be considered in interpreting future observations of giant planet compositions.

\emph{Acknowledgments} 

The authors are grateful to helpful and insightful comments from the anonymous referee on a previous draft of this paper. F.J.C. acknowledges support from NASA’s Emerging Worlds Program, grant 80NSSC20K0333, and Exoplanets Research Program, grant 80NSSC20K0259. M.N.B. thanks A.O. Warren, A.D. Feinstein, C. Keith, and especially B.A. Sanders for their gracious help in debating word choice, critiquing color schemes, and providing all-around emotional support. M.N.B. also thanks the University of Chicago Planet Formation group for their encouragement and insightful questions/comments throughout this project.

\bibliography{references_newactual}

\end{document}